\documentclass[
final            
  ]
  {aipproc}

\usepackage[english]{babel}

\layoutstyle{6x9}

\begin{document}

\title{Techniques to study cosmic ray anisotropies}


\author{Silvia Mollerach}{
  address={CONICET, Centro At\'omico Bariloche, 8400 R\'io Negro, Argentina}
}

\begin{abstract}
These lectures review some of the techniques used to analyse the
arrival direction distribution of Cosmic Rays, and some relevant results 
on the field.

\end{abstract}

\maketitle


\section{Introduction}

The study of the anisotropies in the arrival directions of cosmic rays (CR)
provides a handle, together with the spectrum and composition studies, to face 
several open questions in the field like: which are the sources of cosmic 
rays?, how do CRs propagate?, which is the CR composition?, how are the 
galactic and extragalactic magnetic fields?

The cosmic ray flux has proven to be very close to isotropic and thus
very careful studies are needed to measure the anisotropies.
Before entering the discussion of the techniques used, which are the main 
scope of these lectures, we will review some topics that are relevant to 
understand the kind of anisotropies that can be expected in different
energy ranges. These are: the matter distribution in our neighbourhood,
the galactic and extragalactic magnetic field effects on the CR propagation
(deflections, flux amplification, multiple images) and the Greisen
Zatsepin Kuzmin (GZK) horizon.
  
\subsection{Our local universe}

The distribution of matter around us is not isotropic. 
The Milky Way is a spiral galaxy. The disk, that has a width of about 1 kpc
(1 pc $\sim$ 3 ly), contains in the spiral arms most of the visible 
stars, as well as atomic gas with a density 
of approximately one particle per cm$^3$. The solar system is located 
at 8.5 kpc from the galactic center. A regular magnetic field component 
permeates the disk. The Galaxy is also
permeated by a turbulent magnetic field where charged cosmic ray particles
are trapped and that probably extends few kpc outside the disk.
The Galaxy has also a spheroid (or stellar halo) of more spherical shape 
constituted by an older stellar population and a smaller amount of gas 
(with density $\sim 0.01/$cm$^3$), that extends up to a distance of about
15 kpc. It has a central bulge with a bar-like shape. 
A larger dark matter halo is also present, which existence is
evidenced by its gravitational
effects, such as the rotation curves of stars and gas that remain flat
up to distances larger than those containing the visible matter, 
but its composition is not yet known.
The Galaxy has also a massive black hole in the center with a mass of
few $10^6 $M$_\odot$ .

The Milky Way and its neighbour galaxy Andromeda are the most prominent
members of a small cluster of about 30 galaxies called the Local Group, 
with a diameter of about 2 Mpc.

The Local Group forms part of the Local Supercluster, a large agglomerate 
of about 100 galaxy clusters and groups, that is dominated by the
Virgo cluster located near its center (distant about 18 Mpc from the Milky Way).
Virgo is a large cluster, with more than  2000 galaxies, including the 
prominent radio galaxy M87.

At even larger scales the distribution of clusters shows an
inhomogeneous distribution with large filaments and voids. Only at
scales larger than few hundred Mpc the universe starts to look
homogeneous. 

\subsection{Magnetic field effects on the CR propagation}

The galactic magnetic field has a regular and a turbulent component.
Despite a considerable observational effort, both are still poorly known
\cite{han,beck}.
From the observation of polarized radio emission of other face-on spiral
galaxies it is known that the regular magnetic field follows the
spiral pattern of the stars distribution, while radio polarization 
measurements in edge-on galaxies show the existence of magnetic halos   
extending few kpc above and below the galactic disks.
                              
In our own Galaxy, Faraday rotation measures of pulsars and extragalactic
radio sources indicate that the regular magnetic field follows the spiral 
pattern in the disk, with a local value $B_{reg} \simeq 2 - 3 \ \mu$G. 
\cite{han}. The turbulent field is thought to be larger in rms amplitude than 
the regular one, with a coherence length of approximately 100 pc.

The propagation of charged CRs is affected by the magnetic fields 
(both galactic and extragalactic) present along their trajectories.  
The deflection of ultra-relativistic charged particles is proportional
to their charges and inversely proportional to their energy

$\vec F  =  m  \gamma \dot {\vec v}  =   {q \over c}  \vec v \times \vec B,  
\ \ \ \ \ \ \ \  \vec v =  c  \hat u.$

Then, the direction of propagation is given by  

$\hat u  = \hat u_0 + {Ze \over E} \int dl \ \hat u \times \vec B.$

A CR of charge $Ze$ and energy $E$ in a constant field $B$ describes a
circle of Larmor radius

$r_L   \simeq  {{E/Z} \over {10^{15} eV}}  {{\mu G}\over B}  pc.$          

Then, for $E/Z < 10^{18}$ eV ($\equiv$ 1 EeV) cosmic ray trajectories 
wind around
the magnetic field lines and they remain confined by the galactic magnetic 
field for a long time, while for larger energies their trajectories are
only moderately perturbed and they are not confined in the Galaxy.
For $E/Z < 10^{17}$ eV they scatter off the turbulent magnetic field
irregularities, that have a coherence scale
$l \leq r_L$, make a random walk and diffuse.

\paragraph{Liouville theorem}

An isotropic flux of CRs remains isotropic after
propagating through a magnetic field.
According to the Liouville theorem the phase space distribution
$f(\vec r,\vec p)$ is constant along cosmic ray trajectories if there are no
processes that dis troy or create particles.
The intensity of CRs, defined as the number of particles crossing a
unit area $dA$ per unit time $dt$ and unit solid angle $d\Omega$ with
energy within $E$ and $E+dE$ is given by 
$I = dN/(dA dt d\Omega dE)$ and we can write $dN = f(\vec r,\vec p)
d^3r d^3p$, with $d^3r = dA v dt$ and  $d^3p = p^2 dp d\Omega$.
Then $I = f(\vec r,\vec p) v p^2 dp/dE = f(\vec r,\vec p) p^2$.
Since $p$ is constant along the trajectory, $I$ is also constant.
As a consequence, an isotropic CR flux remains isotropic unless there is a 
'shadowing effect', i.e. there exist directions from which particles
cannot reach the detector coming from infinity.
For example at low energies this happens because of the Earth `shadow':
trajectories of antiparticles leaving from the detector hit the Earth
due to the deflections in the geomagnetic field.
This in particular gives rise to the so-called East-West effect:
Protons with energy smaller than few GeV  are not able to reach the
Earth from the east. The sign of this E-W asymmetry was used to infer
that CR primaries are positively charged.
Towards the poles the threshold is smaller and this has the effect of
increasing the CR intensity with latitude at low energies ('latitude effect').

\paragraph{Deflection of charged particles in the Galactic magnetic field}

\begin{figure}
  \includegraphics[height=.3\textheight]{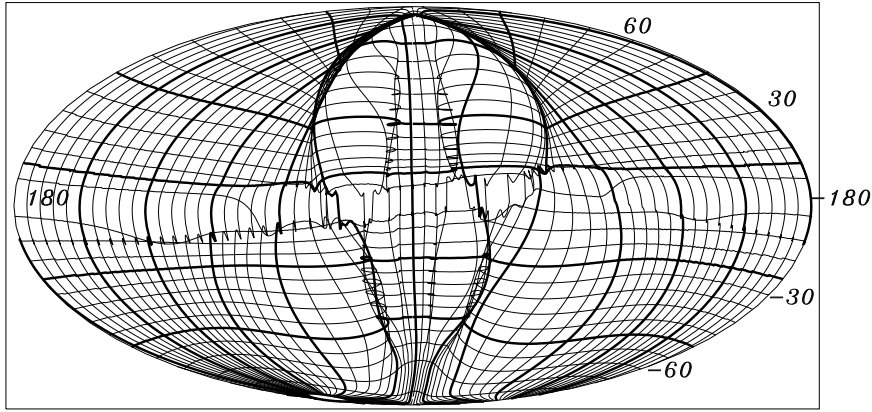}
  \caption{'Sky sheet' in galactic coordinates.}
\label{skysheet}
\end{figure}

As we have discussed the deflection of charged particles is inversely
proportional to their energy, and only for $E/Z \gg 10^{19} eV$ the
deflections in the Galactic magnetic field are expected to become smaller than 
a few degrees and CR astronomy could become feasible.
If the Galactic $B$ field (and composition) were known, one could 
correct the arrival direction to search for the source. This could be
done by `backtracking antiparticles' leaving the detector
with the reverse velocity of the incoming CR.
According to the Liouville theorem the magnetic fields 
cannot produce anisotropies in an isotropic flux, but they
do  in fact affect anisotropic fluxes in different ways: the flux can
be amplified in some regions and deamplified in others in an energy
dependent way, and even multiple images of sources located in some
regions can appear.  
These effects can be visualized plotting for a
regular grid of arrival directions at Earth the corresponding
directions from which the particles arrived to the galactic halo. 
This is shown in Figure \ref{skysheet} for a particular model of the regular
magnetic field and  for particles with $E/Z$ =  20 EeV. 
One may picture this distorted image of the
sky seen from the Earth as a sheet (the `sky sheet') that can be
stretched and folded. A source located in a fold of this sky sheet
will have multiple images, i.e. cosmic rays of the same energy can
arrive to the Earth from several different directions. Moreover, the
flux coming from a source in a region where the sheet is stretched
will appear demagnified while that from a source in a compressed
region will appear magnified \cite{toes}. 

If several CRs with different energies coming from one source are
detected it would be possible to measure the integrated 
perpendicular component of the magnetic field along the CR
trajectory and locate the actual source position.

Magnetic lensing phenomena also appear for turbulent fields \cite{turb}.
The deflections in a turbulent field can be viewed as a 
random walk of the CR particles that suffer in each field domain a
deflection in a random direction. The accumulated deflection after
traversing a distance $L$ in a field with coherence length $L_c$ is
given by 

$\delta_{rms}=\sqrt{L/L_c}(Z e B_{rms} L_c/E)$,

\noindent where the last factor corresponds to the mean deflection in one 
domain.

For $B_{rms} \simeq B_{reg}$, deflections in the turbulent field 
are smaller than those produced by
the regular field, but can dominate the magnetic lensing effects.
Multiple images appear below a critical energy $E_c$, such that 
typical transverse displacements among different paths 
become of order the correlation length of the $B$ field ($\delta_{rms}
\simeq  L_c/L$).
Typically $E_c \simeq 4\times 10^{19} eV\ Z (B_{rms}/5 \mu$G$) (L/2
$kpc$)^{3/2} (L_c/50 $pc$)^{-1/2}$. 
For $E < E_c$ , the number of images grows exponentially.
A regime is reached with a large number of images,
spread over a region of size $\delta_{rms}$  and with mean
magnification $\langle A \rangle \simeq 1$ (like twinkling stars).

Magnetic fields are also present outside galaxies, but the observational 
constraints are still very poor.
The amplitude in the central region of clusters may reach the $\mu$G.
The distribution is believed to follow the filamentary pattern of the
large scale matter distribution.
In most of the space it is usually assumed that 
$B_{rms}  = 10^{-8} - 10^{-9}$ G
and the coherence length  $L_c \sim $Mpc.

\subsection{The Greisen-Zatsepin-Kuzmin horizon}

Soon after the discovery of the cosmic microwave background (CMB) radiation 
it  was realized by Greisen, Zatsepin and Kuzmin \cite{gzk} that the fluxes 
of CR protons with energies of order $10^{20}$ eV and above 
would be strongly attenuated over distances of a few tens of Mpc. This is 
due to the energy losses caused by the photo-pion production
processes in the interactions of the protons with the CMB photons. 
Similarly, if CR sources accelerate heavy nuclei, these can photo-disintegrate 
into lighter ones as they interact with CMB and infrared (IR) photons on 
their journey to us. In this way the fragments may arrive at the Earth with 
significantly smaller energies than the parent nuclei produced at the sources. 
Moreover, both protons and heavy nuclei can further loose energy by pair 
production processes, although due to the small inelasticities involved the
typical attenuation length associated with $e^+e^-$ production at ultra-high 
energies is large ($\sim$ 1 Gpc for protons).
These processes limit the distance from which ultra-high energy cosmic rays
can arrive to the Earth. For example, at energies above $\sim 60$ EeV, 
cosmic rays should mostly come from nearby sources: in the hypothesis of an 
homogeneous distribution of sources, $90\%$ of the flux should come
from sources 
closer  than $\sim 200$  Mpc and  $50\%$ from sources closer than $\sim 100$ 
Mpc. The distance limits turn out to be similar for iron nuclei (although the
responsible processes are different), while lighter nuclei should come from a
much closer neighbourhood \cite{horizon}.

\subsubsection{Exposure}

For any CR anisotropy analysis it is very important to have a good estimate
of the expected flux in any direction for the given experimental 
setup in the case the CR flux is isotropic.
The exposure measures the time integrated effective collecting area 
in units of km$^2$ yr. For each direction of the sky $\omega(\delta,\alpha)$
gives the relative exposure. For a detector in continuous operation it is 
uniform in right ascension $\alpha$, being only a function of the declination 
$\delta$.

\begin{figure}
  \includegraphics[height=.3\textheight,angle=270]{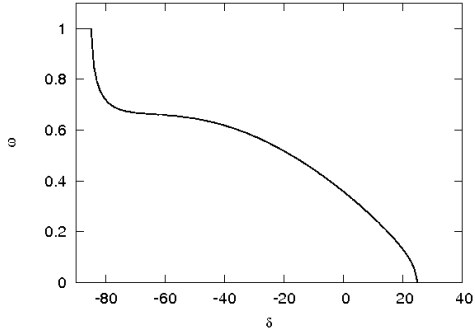}
  \caption{Exposure as a function of declination}
\label{expo}
\end{figure}

If the detector is fully efficient for particles arriving with zenith angle
$\theta < \theta_m$, the exposure has only a $\cos(\theta)$ modulation due 
to the change in the effective detection area. The zenith of a detector 
located at latitude $\delta_0$ corresponds to a position in the celestial 
sphere given $\hat \xi = (\cos \delta_0 \cos \alpha_\xi, \cos \delta_0 \sin 
\alpha_\xi, \sin \delta_0)$, where $\alpha_\xi$ is the right ascension of the 
detector zenith. A source in a direction $\hat s = (\cos \delta_s \cos 
\alpha_s, \cos \delta_s \sin \alpha_s, \sin \delta_s)$ is seen at a zenithal 
angle $\cos \theta = \hat \xi \cdot \hat s = \cos \delta_0 \cos \delta_s 
\cos(\alpha_s-\alpha_\xi) + \sin \delta_0 \sin \delta_s$.
The exposure towards a direction $\hat s$ is proportional to the integral of 
$\cos \theta(t)$ for all the times when $\theta(t) < \theta_m$, and is given 
by $\omega(\delta) \propto \cos \delta \cos \delta_0 \sin \alpha_m +
\alpha_m \sin \delta \sin \delta_0$, where $\alpha_m = \arccos(z)$ (if
$-1 < z < 1$) with $z = (\cos \alpha_m - \sin \delta_0 \sin \delta)/\cos 
\delta \cos \delta_0$,  $\alpha_m = \pi$ (if $z < -1$) and $\alpha_m = 0$ 
(if $z > 1$) \cite{sommers}. 
The exposure is shown in Figure \ref{expo} for a detector at latitude 
$-35^\circ$, corresponding to the Pierre Auger Observatory.

Perfect exposure holds for a continuously operating detector
at the highest energies where every shower triggers
the detector. At lower energies the more inclined showers are more attenuated 
than the vertical ones and are less effective to trigger the detector, and 
then the exposure is no longer just proportional to $\cos \theta$. 
There are two main strategies to compute the exposure in these cases:
a semi-analytical method that consists in obtaining the zenith angle 
distribution from the data itself and then proceed as before with the 
integration in $\theta(t)$, or the shuffling technique that consists in 
simulating a large number of fake events using the time and zenith angle 
distribution of real events from which the isotropic expectation can
be obtained.

\section{Cosmic ray anisotropies}

Different signals of anisotropies are expected to appear at different energies
due to the different sources and propagation effects involved that we will 
briefly discuss here.
  
At the highest energies the magnetic deflections are expected to be small 
and the GZK horizon limits the distance from which CR can arrive, then we 
might expect to observe only events coming from 'nearby' sources.
The expected signals are then: small scale clustering of events coming from 
the same source, correlation of events with a population of source candidates,
intermediate scale clustering reflecting the clustering of local sources. 
Lowering the energy, the deflections increase and the GZK horizon also
increases,
then the CR flux is expected to become more isotropic, but we can still expect 
some intermediate scale clustering and correlation with the sources 
distribution. Although the distribution is expected to become more isotropic
when lowering the energy, the increase in statistics should help to detect the
smaller anisotropy signal. At even lower energies, a large scale 
anisotropy signal coming from the diffusion and drifts of CR in the Galactic 
magnetic field is expected to be seen.
  
If there is a neutral component of CRs a point-like signal is expected and a 
correlation with the source population at the angular resolution scale should
appear.

In addition to these intrinsic anisotropies in the CR arrival direction, there
is also a large scale anisotropy signal expected from the motion of the 
detector with respect to the CR rest frame, the so-called Compton
Getting effect. 

We will now discuss some of the techniques that are used to measure the 
anisotropies in the CR intensity $I$, that is defined as the number of 
particles per unit solid angle that pass per unit time through a unit of area 
perpendicular to the direction of observation $\hat u$. The differential 
(spectral) intensity $I(E)$ is the intensity of particles with energy in the 
interval from $E$ to $E+dE$. 

\subsubsection{Large scale anisotropies}

The first signal at large angular scales that we look for is a dipole in some
direction $\hat j$. A dipole gives rise to an intensity 
$I(\hat u) = I_0 + I_1 \hat j \cdot \hat u$. 
The amplitude is defined as $\Delta = 
(I_{max}-I_{min})/(I_{max}+I_{min})= I_1/I_0$.

The most usual methods to detect a dipole are one dimensional: they only 
study the right ascension distribution (of the full data set or in a fixed 
declination band). The reason is that some experiments cannot reliably 
determine the dependence of the exposure in the declination.

The most standard analysis technique is the Rayleigh method, that performs an 
harmonic analysis in right ascension \cite{linsley}. For $N$ events
with right ascension 
$\alpha_i$, the k-order harmonic has amplitude $r_k = \sqrt{a_k^2+b_k^2}$ and 
phase
$\phi_k = \arctan(b_k/a_k)$, with $a_k = (2/N)\sum_{i=1}^N \cos(k \alpha_i)$
and $b_k = (2/N)\sum_{i=1}^N \sin(k \alpha_i)$.

The significance of a given $r_k$ measurement, that is the
probability that an amplitude larger or equal than the observed $r_k$ 
arises from an isotropic data set by chance, can be estimated by
$P(\geq r_k) = \exp (-N r_k^2/4)$. According to the Central Limit Theorem,
the sum of $N$ independent random variables $x_1,\dots, x_N$ identically
distributed with any probability distribution function (pdf) 
has a pdf approaching for large $N$ a Gaussian with mean equal to the sum of 
the means and variance equal to the sum of the variances. Taking as the $x_i$ 
variables the right ascension coordinates of the events $\alpha_i$, that have 
a uniform distribution in the interval $[0,2\pi]$ for an isotropic distribution
of CRs, we find that both $a_k$ and $b_k$ are Gaussian distributed with
$\langle a_k \rangle = \langle b_k \rangle = 0$ and $\sigma^2(a_k) = 
\sigma^2(b_k) = 2/N$. To obtain the distribution of the amplitudes $r_k$, we 
have to consider that for $n$ Gaussian variables $y_1,\dots,y_n$, the variable
$z = \sum_i(y_i-\langle y_i \rangle)^2/\sigma_i^2$ has a $\chi^2$ distribution 
with $n$ degrees of freedom. Then, the variable $z$ defined as
$z = (N/2) (a_k^2 + b_k^2) = (N/2) r_k^2$ has a $\chi^2(2)$ distribution,
$P(z) = \exp(-z/2)/2$. Changing variable to $r_k$ we get $P(r_k) d r_k = 
(1/2) \exp(-N r_k^2/4) N r_k d r_k$, and integrating it above a given amplitude
we get the advertised $P(\geq r_k)$.

The Rayleigh analysis only has information on the projection of the real dipole
into the equatorial plane. For a full sky uniform exposure experiment, $r_1 = 
\Delta \cos(\delta_{dip})$, with $\delta_{dip}$ the dipole direction
declination, and $\phi_1$ is the right ascension of the dipole 
direction.

If the exposure is not uniform or there is no full sky coverage,
the relation between
$r_1$ and the original dipole components 
$\Delta_z\equiv \Delta\sin\delta_{dip}$ and 
$\Delta_\perp\equiv \Delta\cos\delta_{dip}$ in the case in which the exposure is 
independent of $\alpha$,
is given by \cite{ap}

$$r_1=\left|{c_3 \Delta_\perp\over c_1+c_2 \Delta_z}\right|$$
where

$c_1=\int_{\delta_{min}}^{\delta_{max}}{\rm d}\delta\ \omega(\delta)
  \cos\delta,\ \ \ \ 
c_2=\int_{\delta_{min}}^{\delta_{max}}{\rm d}\delta\ \omega(\delta)
  \cos\delta \sin\delta, \ \ \ \ 
c_3=\int_{\delta_{min}}^{\delta_{max}}{\rm d}\delta\ \omega(\delta)
  \cos^2\delta. $

Some proposals to reconstruct the three dimensional dipole can be found in refs.  \cite{ap,mr}.

\paragraph{Compton Getting effect}

If the CR flux is isotropic in a reference system $S$ and the observer 
is moving with respect to that coordinate system with a velocity $\vec V$,
he will measure a dipolar anisotropic flux.
Let $f(\vec p,\vec r)$ be the distribution function of CR particles in
the frame $S$, where it is isotropic, and  $f'(\vec p',\vec r')$ that in
$S'$, that corresponds to the detector frame moving with $\vec V$ with
respect to $S$. Due to Lorentz invariance $f(\vec p,\vec r) = f'(\vec
p',\vec r')$. The momentum of particles in $S'$ is related to that in
$S$ by $\vec p'= \gamma_V (\vec p -(p/u) \vec V)$ with $u$ the
velocity of the relativistic particles. For a non-relativistic motion
of the detector we take $V \ll c$ and $\gamma_V \sim 1$. Then, we can
write 

$f'(\vec p') = f(\vec p') - \frac{\partial f}{\partial \vec p'}
\cdot \vec V \frac{p}{u} = f (1 - \frac{\vec V \cdot \vec p}{u
p}  \frac{\partial \ln f}{\partial \ln p}).$

The intensity can be written as $I (t,E,\vec r,\hat n) = p^2 f(t,\vec
r,\hat p)$, then  $\ln I = 2 \ln p - \ln f$ and $\partial \ln
f/\partial \ln p =  \partial \ln I/\partial \ln p - 2 \simeq
(\partial \ln E/\partial \ln p)(\partial \ln I/\partial \ln E) - 2 =
(1 - m^2/E^2)(-\gamma) - 2 \simeq -(\gamma +2)$ for particles with
spectrum $I \propto E^{-\gamma}$. Then
$$I' (E') = I \big(1 + \frac{V}{u} (\gamma +2) \cos \theta\big).$$
For example for a detector moving with $V = 100$ km/s and $\gamma \sim
3$ the dipole amplitude is $\Delta \sim 1.6 \times 10^{-3}$.

The orbital motion of the Earth around the Sun is expected to modulate the 
measured flux of CRs due to the Compton Getting effect. 
The rotation velocity of the Earth around the Sun is $V = 29.8$ km/s.
A vertically looking detector should see a modulation of the intensity 
with the solar time
$I(t) = I_0 (1 + r \cos ((t - t_0) 2 \pi/24 hs))$. For every detector the
maximum appears at a solar time $t_0 = 6$ hs, 
this can easily be understood by thinking about the relative direction 
of the zenith and the rotation of the Earth.
The amplitude depends on the detector's latitude and can be as large as
$\Delta = (V/c) (\gamma + 2) \simeq 5 \times 10^{-3}$.
A modulation in the solar time frequency that agrees with that expected from 
the Compton Getting effect has in fact been measured  at energies around 10 
TeV by the EAS-TOP experiment \cite{aglietta96} and by the Tibet Air-Shower 
experiment \cite{amenomori04}. The measured amplitude of the solar frequency 
modulation has also been used to estimate the spectral index, obtaining 
$\gamma = 3.03 \pm 0.55$ at an energy range (6 - 40) TeV \cite{amenomori07},
in good agreement with the direct measurement value.

\paragraph{Large scale anisotropy measurements}

The good agreement with the expectations for the solar frequency measurements, 
where the expectations are well known, gives confidence that the measurements 
are reliable for the sidereal frequency analysis, that contains the real right
ascension modulation of the CR intensity, for which the expectations are
uncertain.

A contribution to the sidereal time frequency (or right ascension) modulation
is also expected from the (unknown) motion of the solar system with respect 
to the rest frame of the CRs.
(The solar day is a bit longer than the sidereal day: 1 year = 365.24 solar 
days = 366.24 sidereal days). 
\begin{figure}
  \includegraphics[height=.3\textheight]{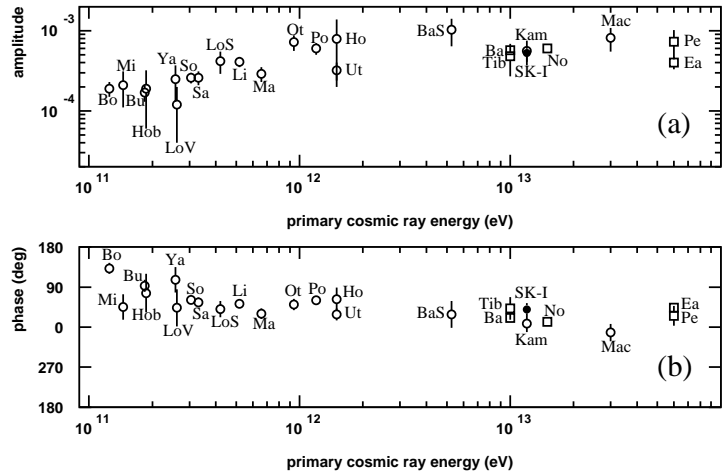}
  \caption{Amplitude and phase as a function of the energy from
    different CR experiments \cite{guillian07}.}
\label{ra}
\end{figure}

The results of the sidereal harmonic analysis by different experiments 
are summarized in Figure \ref{ra} \cite{guillian07}. 
The measured amplitudes in the energy range 
from $10^{11}$ to $10^{14}$ eV are $\Delta \simeq$ few $\times 10^{-4}$.
The sidereal time modulation arises from a combination of the intrinsic
anisotropy of the CR intensity in their own frame and the contribution
from the detector motion. If the CR plasma were at rest with respect 
to an inertial system  attached to the Galactic center, the rotational 
velocity of the Sun around the Galaxy of $\sim 220$ km/s, would
lead to a dipole amplitude $\Delta \simeq$ few $\times 10^{-3}$, an order 
of magnitude larger than the observed values, indicating that the CR plasma
corrotate with the local stars \cite{amenomori06}.

\begin{figure}
  \includegraphics[height=.3\textheight]{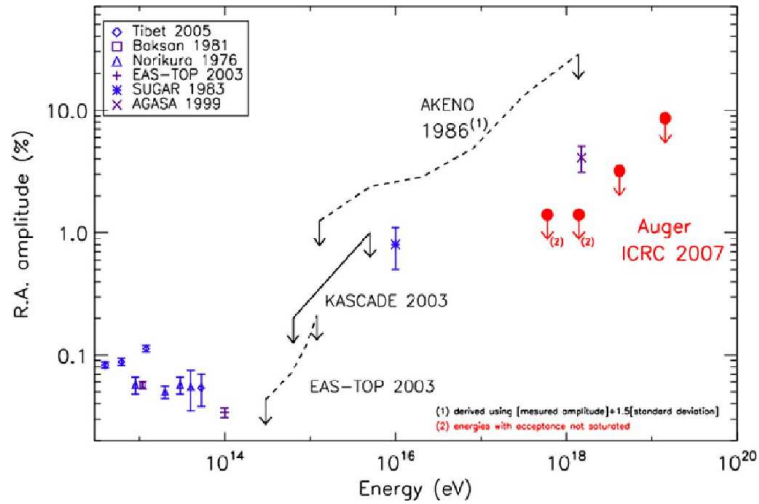}
  \caption{Summary of Auger 95\% CL upper bounds on the 
amplitude of a
  dipolar  modulation in right ascension and results from previous
  experiments.\cite{armengaud}}
\label{hedip}
\end{figure}

The transport of galactic CRs in the magnetized plasma is governed by 
anisotropic diffusion, drift and convection and detailed measurements
of CR anisotropies can be useful to explore magnetic field 
characteristics and the CRs transport. In particular, precise measurements
of the anisotropies in the knee of the spectrum may be very helpful to 
understand the origin of this feature. If the knee is due to the limit in 
the acceleration power of galactic sources, a decrease in the dipole 
amplitude with increasing energy is expected as the more isotropic 
extragalactic component enters in the play. Instead, if the knee is due to
the fact that as the energy increase Galactic CRs start escaping more 
easily from the Galactic magnetic field, then an increment of the 
amplitude is expected with increasing energy as CRs flow more
efficiently out of the Galaxy.
Experimental measurements of the anisotropies cannot yet settle this
point. The results at the knee and higher 
energies by different experiments are summarized in Figure \ref{hedip} 
\cite{armengaud}.

As the energy increases and the number of events diminishes it becomes 
more difficult to measure the dipole amplitude.
At energies higher than $10^{18}$ eV AGASA has found a Rayleigh
amplitude of $4\%$. More recent results from the Pierre Auger Observatory
have put upper bounds of $1.4\%$ above $10^{18}$ eV and $10\%$ above 
$10^{19}$ eV as it is shown in Figure \ref{hedip}.

\subsubsection{Small and intermediate scale clustering}

Clustering at small scales can be the clue to detect repeating sources. 
The amount of clustering (or the fraction of repeaters) gives a 
measure of the number of sources that contribute to the CRs above a
given energy 
threshold, from which we can deduce the local density of sources.
Clustering at intermediate angular scales contains information on the
pattern of the distribution of the local sources.

\paragraph{The autocorrelation function}

The autocorrelation function is a standard technique to analyse the
distribution of points in the sky. It measures the excess (deficit)
in the number of pairs with respect to that expected from an isotropic 
distribution as a function of the angle.
For an isotropic distribution of $N$ points on the full celestial
sphere the expected number of pairs with angle smaller than $\alpha$ 
is given by $n_p = (N (N-1)/2) (1- \cos(\alpha))$.
For partial/non-uniform sky coverage the expected number of pairs
from an isotropic flux has to be computed simulating isotropic event
realizations following the exposure and counting the pairs as 
a function of the angle. 
The number of pairs separated by less than an angle $\alpha$ among the 
$N$ events with energy larger than a given threshold $E$ is
\begin{equation}
n_p(\alpha)=\sum_{i=2}^{N} \sum_{j=1}^{i-1} \Theta(\alpha-\alpha_{ij}),
\end{equation}
where $\alpha_{ij}$ is the angular separation between events $i$ and $j$ and
$\Theta$ is the step function.
The chance probability for any excess of pairs at
a fixed angle $\alpha$ and energy threshold is found from 
the fraction of simulations with a larger or equal number of 
pairs than what is found in the data at the angular 
scale of interest.

The results of the autocorrelation function analysis depend on the
chosen values of $\alpha$ and $E$ (with the corresponding number of events 
$N$ above that energy threshold). 
The fact that the deflections expected from galactic and extragalactic 
magnetic fields and the distribution of the sources are largely unknown
prevents to determine these values a priori.  The significance 
of an autocorrelation signal at a given angle and energy, when 
these values have not been fixed a priori, is a delicate issue.
A possible solution to this problem was proposed by Finley and Westerhoff 
\cite{FW}, in which a scan over the energy threshold and the angular 
separation is performed. For each value of $N$ and $\alpha$, the fraction 
$f$ of simulations having an equal or larger number of pairs than the data 
is computed. The most relevant clustering signal corresponds to the values 
of $\alpha$ and $N$ that have the smallest value of $f$, referred to as 
$f_{min}$. To establish the statistical significance of a given excess, 
it is necessary to account for the fact that the angular bins, as well as 
the energy ones, are not independent. This can be done performing a large 
number of isotropic simulations with the same number of events as the data 
and calculate for each realisation the most significant deviation $f_{min}^i$. 
The statistical significance of the deviation from isotropy is the integral 
of the normalised $f_{min}$ distribution above $f_{min}^{data}$. 
Then the probability that such clustering arises by chance from an isotropic 
distribution can be estimated just from the fraction of simulations having 
$f_{min}^i \le f_{min}^{data}$.

Although the data from a number of experiments
have shown a remarkably isotropic distribution of arrival directions, there
has been a claim of small scale clustering at energies larger than 
$40$ EeV by the AGASA experiment \cite{AGASA}. The most recently published
analysis \cite{teshima} reports 8 pairs (five doublets and a triplet) with 
separation smaller than $2.5^\circ$ among the 
59 events with energy above 40 EeV, while 1.7 were expected from an isotropic 
flux. The probability for this excess to happen by chance 
was estimated to be less than $10^{-4}$. The significance of 
the AGASA clustering result was, however, subject of debate based on the 
concern that the energy threshold and angular separation were not fixed 
a priori. Tinyakov and Tkachev \cite{TT01} computed the penalisation arising
from making a scan in the energy threshold and obtained a probability 
of $3\times 10^{-4}$. Finley and Westerhoff \cite{FW} took also into account 
the penalisation for a scan in the angular scale and obtained a probability
of $3.5\times 10^{-3}$. The HiRes observatory has found no significant 
clustering signal at any angular scale up to $5^\circ$ for any energy 
threshold above $10$ EeV \cite{hires}. A hint of correlation at scales around 
$25^\circ$ and energies above $40$ EeV, combining data from HiRes stereo, 
AGASA, Yakutsk and SUGAR experiments has been pointed out in ref.
~\cite{KS}.

\begin{figure}
  \includegraphics[height=.3\textheight]{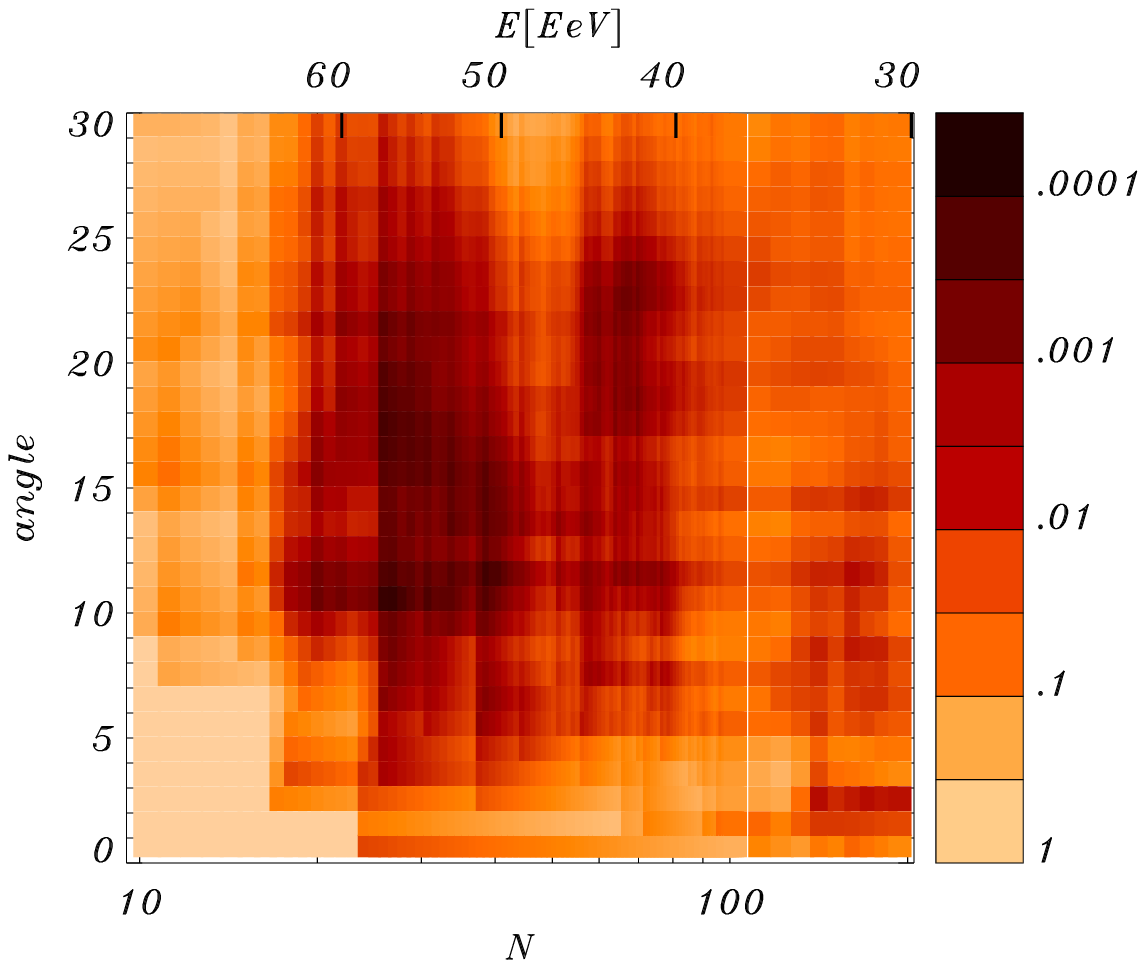}
\includegraphics[width=3in]{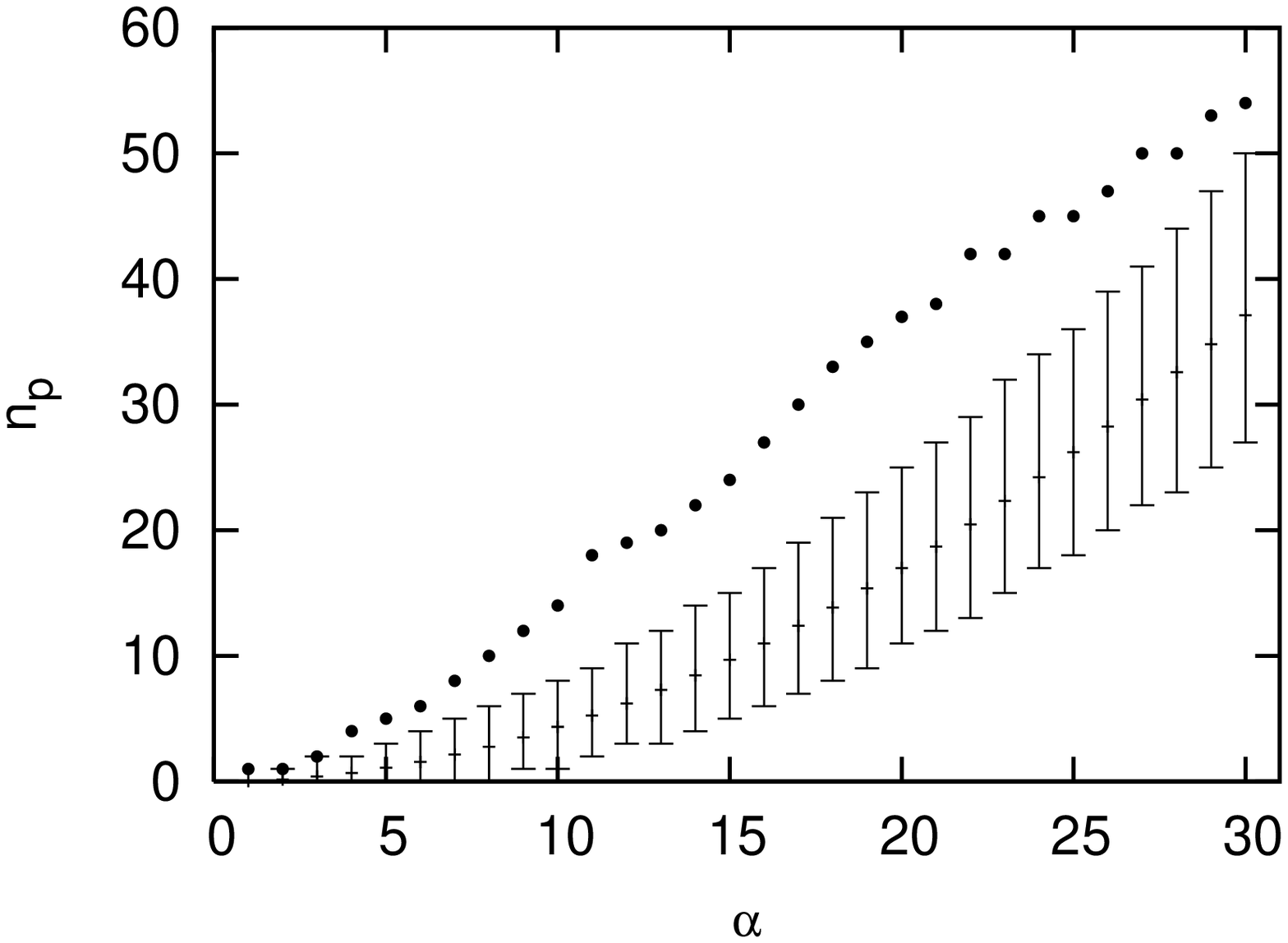}
  \caption{Autocorrelation scan}
\label{scan}
\end{figure}

The Auger collaboration, using the events recorded by the surface detector 
between January 2004 and August 2007, has recently reported the results
of the scan performed in energy (above 30 EeV, with 203 events detected) and 
angular separation (between 1$^\circ$ and 30$^\circ$) \cite{cris08}. 
The results are
shown in the left panel of Figure \ref{scan}. 
The most significant excess of pairs found is for $E > 57$ EeV 
(corresponding to 27 events) and 
$\alpha = 11^\circ$,  with a  chance probability of $P = 1.6 \times 10^{-2}$
to arise from an isotropic distribution. 
Above this energy, a broad region of low probabilities is observed, 
for angular scales between $9^\circ$ and $22^\circ$. The right panel
of Figure \ref{scan} shows the number of observed pairs as a function
of the angle (dots) as well as the expectations from an isotropic
distribution with the 90$\%$ CL bars. 

\paragraph{Search for point-like or extended excesses of events around 
a given direction in the sky}

A neutral component of cosmic rays or a powerful source of charged particles 
at high energies could lead to a point-like excess of CRs arriving from the 
source direction. At lower energies charged particles can give rise to an
extended excess of events from a region close to the source.
Then it is important to have the tools to identify and estimate the 
significance of an excess of events from some region. The events coming from a
source will be superimposed with that from the background, so we need to make an
estimate of the signal and its significance.

For any given direction the first step is to measure the observed number of 
events in a window (that can be a top-hat, Gaussian, etc) around the given 
direction. For a point-like excess (as could arise for neutral primaries) 
the angular resolution size is considered. We call this number $N_{ON}$.
Then we need to estimate the background. For this scope we can use the 
detector measurements in other regions of the sky, $N_B = \alpha N_{OFF}$
with $N_{OFF}$ the events measured in the $OFF$ region and $\alpha = 
t_{ON}/t_{OFF}$ = $ \omega_{ON}/\omega_{OFF}$. From this we can estimate 
the signal as $N_S = N_{ON} -\alpha N_{OFF}$. 
 To estimate the significance, one possibility is to use the variance of 
the signal. As $N_{ON}$ and $N_{OFF}$ are independent measurements,
$\sigma^2(N_S) = \sigma^2(N_{ON})+ \alpha^2  \sigma^2(N_{OFF}).$
Then there are different possibilities to estimate the variances.
One possibility is to just consider two Poisson processes, then
$\hat \sigma_1 (N_S) = \sqrt{N_{ON} + \alpha^2 N_{OFF}}$
 and the significance of the excess is given by 
$S_1 = (N_{ON}-\alpha N_{OFF})/\sigma_1 (N_S)$.
Another possibility is to consider that to estimate the significance of 
an excess what we want is to asses the probability that it arises only 
from the background. Then we would take for the pdf of $N_{ON}$ a Poisson
distribution with mean equal to $\langle N_B \rangle$ and for $N_{OFF}$ a 
Poisson distribution with mean $\langle N_B \rangle /\alpha$. Then the 
variance is given by $\sigma^2(N_S) = \langle N_B \rangle (1 + \alpha)$ and
estimating $\langle N_B \rangle= t_{ON} (N_{ON}+N_{OFF})/(t_{ON}+t_{OFF})=
(N_{ON}+N_{OFF}) \alpha/(1+\alpha)$, we have  
$\hat \sigma_2 (N_S) = \sqrt{\alpha (N_{ON} + N_{OFF})}$ and
$S_2 = (N_{ON}-\alpha N_{OFF})/\sigma_2 (N_S)$.
An observed excess $N_S$ can be said to be an 'S standard deviation detection'.
If $N_{ON}$ and  $N_{OFF}$ are not too low ($\geq 10$), under the assumption 
that all the events come from the background ($\langle N_S \rangle =0$), the
distribution of $S$ approximates a Gaussian variable with zero mean and 
unit variance, and the Gaussian probability of $S$ can be taken as the 
confidence level of the observational result. Tests with numerical 
simulations indicates that $S_2$ is a better estimator than $S_1$ \cite{lima}.
 
Another proposal to estimate the significance is to use a likelihood ratio
method (Li-Ma \cite{lima}). The likelihood ratio of the 'null hypothesis',
corresponding here to no source ($\langle N_S \rangle =0$), 
and that all events are coming from the background, 
and the alternative tested hypothesis, corresponding
to a non-vanishing source ($\langle N_S \rangle \neq 0$), is defined as

$\lambda = \frac{L(data | null \ hyp)}{L(data | alternative \ hyp)} =
\frac{P(data | \langle N_S \rangle =0)}
{P(data | \langle N_S \rangle \neq 0)}.$

If the null hypothesis is true and $ N_{ON}, N_{OFF} \geq 10$, then 
$\sqrt{- 2 \ln \lambda}$ is Gaussian distributed with $\sigma^2 = 1$.
Writting

$P(data | \langle N_S \rangle =0) = P(N_{ON}, N_{OFF} | \langle N_S 
\rangle = 0, \langle N_B \rangle = \frac{\alpha}{1+\alpha} (N_{ON} +
N_{OFF})) = $
 
$\ \ \ \ \ \ \ \ \ \ \ \ \ \ \ Poisson(N_{ON}; \langle N_{ON} \rangle 
=  \langle N_B \rangle)
\ Poisson(N_{OFF}; \langle N_{OFF} \rangle =  \langle N_B \rangle /\alpha ),$

$P(data | \langle N_S \rangle = N_{ON} - \alpha  N_{OFF}, \langle N_B \rangle 
= \alpha N_{OFF}) = $

$\ \ \ \ \ \ \ \ \ \ \ \ \ \ \ Poisson(N_{ON}; \langle N_{ON} \rangle = N_{ON})
\ Poisson(N_{OFF}; \langle N_{OFF}
\rangle = N_{OFF}),$

we can compute $\lambda$ as

$\lambda = \bigg[\frac{\alpha}{1+\alpha}\bigg(\frac{N_{ON} + N_{OFF}}{N_{ON}}
\bigg)\bigg]^{N_{ON}}    
\bigg[\frac{1}{1+\alpha}\bigg(\frac{N_{ON} + N_{OFF}}{N_{OFF}}
\bigg)\bigg]^{N_{OFF}}.$

The Li-Ma significance $S = \sqrt{-2 \ln \lambda}$ was shown to follow 
a Gaussian distribution better that $S_1$ and $S_2$ \cite{lima}.

The search of excesses of flux from point-like and extended regions is
common to both the Cosmic Ray and Gamma Ray astronomy fields. The selection of
the $ON$ and $OFF$ is decided according to the observational data.
For example, if a possible excess of cosmic rays from a predetermined 
direction in sky is tested, the rest of the sky can be used as the 
$OFF$ region to determine the background isotropic expectation and the significance.
If a blind search of excesses at a given angular scale is performed in the
whole sky, proceeding in the same way for each particular direction, a 
distribution of significances is obtained. If no sources are present, they
follow a Gaussian distribution with unit variance.  

\paragraph{Galactic center searches}

\begin{figure}
  \includegraphics[height=.3\textheight]{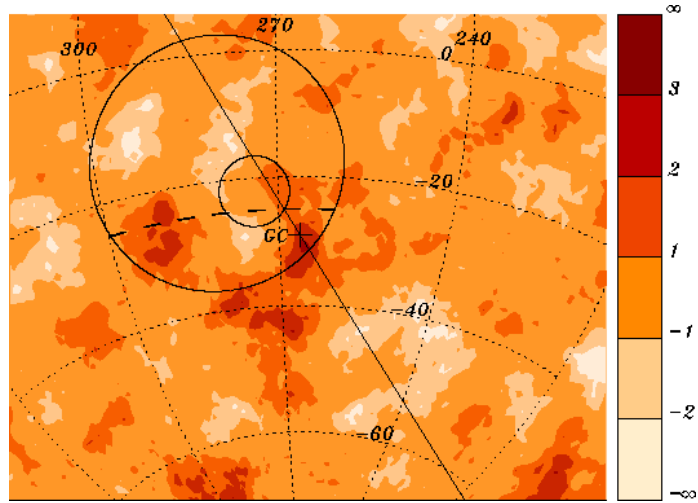}
  \caption{Map in equatorial coordinates  around
  the GC (cross) showing the significance of the
  overdensities in $5^\circ$ radius windows, for $10^{17.9}
  {\rm eV}<E<10^{18.5}$~eV. }
\label{gc}
\end{figure}

A potential site for the acceleration of CRs in our Galaxy is
the Galactic center region with its super massive black hole and high 
density of stars. Thus, it is an interesting region to look for an excess 
of events. The AGASA collaboration found a 4.5$\sigma$
excess ($observed/expected=506/413.6$) in a $20^\circ$ radius region
close to the Galactic center region for the energy range 
$10^{18}$--$10^{18.4}$~eV.
Being in the southern hemisphere, the Auger Observatory has a
privileged view towards the galactic centre (GC), 
which passes at just $6^\circ$ from the zenith at the site. 
For the same region where AGASA found the excess, and the same
energy range, Auger data led to $obs/exp=2116/2159.6$ \cite{gcenter}, a result 
 inconsistent with a large
excess. Similarly, an excess reported by the SUGAR collaboration in a $5^\circ$
region slightly displaced from the GC  was not confirmed
by Auger. A map of
overdensity significances  on $5^\circ$ radius windows in the region around
the GC is shown in fig. \ref{gc}, together with the regions were the
AGASA and SUGAR excesses were reported. The excesses present
in this map 
are consistent with the expectations from fluctuations of an isotropic
distribution. 

\subsubsection{Searches for correlations with objects}

Given a population of candidate sources there are different proposed tests
to search for a correlation with CR arrival directions.

\paragraph{Cross-correlation function}

This technique 
looks for an excess of CR separated by less than a given angle from
any candidate source in the set with respect to the expectations from an 
isotropic CR distribution. The procedure is very similar to the 
autocorrelation analysis: we first count the number of pairs CR-objects as a 
function of the angle in the data. Then, we repeat the procedure for a large 
number of isotropic simulated data sets. To estimate the significance of any 
excess we compute the fraction of the simulations with larger number of pairs
than those present in the data. This method was for example used to look for
a possible correlation of CR with $E > 10$ EeV and BL Lacs with magnitude 
$m < 18$ at the experimental angular resolution scale (first found in HiRes 
data by Gorbunov et al. \cite{gor2004}). In that magnitude range there are 
156 BL Lacs in the field of view of HiRes in the Veron Cetty and Veron 
catalogue \cite{VC} and 271 events have been reported by HiRes in that energy 
range. The number of observed pairs within $0.8^\circ$ was 11, while only 
3 were expected from an isotropic distribution. The fraction of isotropic 
simulations with a larger number of pairs is $f (0.8^\circ) = 4 \times 
10^{-4}$. The penalization for searching at different angles and with 
different sets of objects is not included in that figure.
A test of the signal using Auger data, with 1736 events with $E > 10$ EeV
showed no evidence of excess of correlation with BL Lacs with the same 
magnitude limit in the same catalogue \cite{harari2007}.

\paragraph{Maximum likelihood ratio method}

The idea in this method is, for a given model of the source distribution, to
find out the values of the parameters that lead to a better agreement with
the observational data. 
For example, if the model considered is that a fraction of 
CRs events come from a known population of sources and another fraction 
from an isotropically distributed background, we would say that from the set
of $N$ measured events, $n_s$ are source events and $N - n_s$ are
background events. The probability that a background event arrives from 
any direction is proportional to the exposure $\omega (\hat u)$. 
The probability that an event comes from one of the sources is peaked 
around the source direction $\hat s$ with some window $Q(\hat u,\hat s)$
given by the angular resolution of the experiment (for charged particles a 
larger spread can be introduced to account for magnetic deflections).
For $M$ sources, the contribution of all the sources weighted by the
exposure and eventually by a relative source intensity is added. 
For the simplest option of equally apparent bright sources we have
$$Q(\hat u) = \sum_j \omega (\hat s_j)Q(\hat u, \hat s_j)/\sum_j \omega 
(\hat s_j).$$
The probability distribution for any event is
$P(\hat u) = (n_s/N) Q(\hat u) + (N-n_s)/N \omega (\hat u).$
The likelihood for the set of $N$ events is 
$L (n_s) = \prod_{i=1}^N P(\hat u_i).$
Then we search for the value of $n_s$ that maximizes the ratio 
$R(n_s) = L(n_s)/L(0)$, with $L(0)$ the likelihood of the null hypothesis 
($n_s = 0$). The significance can be estimated by computing $R$ for a large
set of isotropic simulations and counting the fraction with $R_{sim} \ge
R_{dat}$.

This method was used by HiRes to test the correlation of CRs with $E > 10$ EeV 
with BL Lacs with $m < 18$ at the resolution angular scale in their data 
\cite{Abbasi2005}.
For each event $i$ they used for $Q(\hat u_i,\hat s_j)$ a Gaussian centred at 
$s_j$ with a dispersion equal to the angular resolution of that event.
They found that $\ln R$ is maximized for $n_s = 8.0$ corresponding to 
$\ln R = 6.08$. The fraction of simulations with higher $\ln R$ is $f = 2 
\times 10^{-4}$. Thus, the results are similar in this case to those
using the cross-correlation analysis.
Some advantages of this method are that it can be adapted to give 
different weight to each candidate source,
for example depending on the distance or known brightness in some
band. It is also possible to consider different angular scales depending
on the angular resolution of the event, or e.g. from the expected magnetic 
deflections in different directions.

\paragraph{Binomial probability scan}

For a given candidate source population, e. g. AGNs, galaxy clusters,
radio galaxies, there are different parameters that will influence
the correlation with events but that are difficult to fix a priori:
the angular scale (magnetic deflections are not known), maximum distance to 
the objects (UHECR from distant sources will have their energy diminished 
by interactions with CMB through the GZK effect), energy threshold (only 
high energy events are expected to be correlated with local sources).
The idea behind this method is to scan in the unknown parameters. For a 
given candidate source population, we can estimate the probability that
an individual event from an isotropic flux has an arrival direction closer
than some particular angular distance $\Psi$ from a member of the catalog,
$p$, by computing the exposure-weighted fraction of the sky which is covered 
by windows of radius $\Psi$ centered on the selected objects. This will be 
a function of angular scale  $\Psi$ and the maximum distance to the objects
considered $D_{max}$, $p(\Psi, D_{max})$. For each energy threshold $E_{min}$, 
with $N$ events over the threshold, the number of events $k$ correlated to the
sources is calculated. Then, the probability $P$ that $k$ or more events of 
the total of $N$ events are correlated by chance with the selected objects 
is given by the cumulative binomial probability
$$P = \sum_{j=k}^N \left(\matrix{N\cr j}\right) p^{\/j}(1-p)^{N-j}~.$$
The more significant correlation in the data set corresponds to the values
$\Psi$, $E_{min}$ and $D_{max}$ that give rise to the smaller $P$ value,
$P_{min}$.
The significance of a given correlation can be estimated performing a large 
number of isotropic simulations and under the same scan in $\Psi$, $E_{min}$ 
and $D_{max}$ obtain the fraction $f$ having a $P_{min}$ smaller than the data.

\paragraph{Correlation of UHECR with AGNs}

To test for possible correlations with  extragalactic 
sources the Auger collaboration analyzed the arrival directions of the events 
above $4\times 10^{19}$~eV to look for coincidences with the positions of the
known  nearby (less than 100~Mpc) active galactic nuclei from the
Veron-Cetty and Veron catalog 'cite{VC}.   The results of a scan over the 
angle $\psi$ between the events and the AGNs, 
the maximum AGN redshift considered $z_{max}$  and the threshold energy
$E_{th}$ show a deep minimum in the probability $P$ of observing a similar
or larger number of correlations arising from isotropic simulated data. This
minimum is obtained for $\psi=3.2^\circ$, $z_{max}=0.017$ (or maximum AGN
distance of $71$~Mpc) and $E_{th}=57$~EeV (corresponding to
the 27 highest energy events) \cite{science,longagn}. Only $\sim
10^{-5}$ of the isotropic simulations have a deeper minimum under a 
similar scan.
In particular, for these 27 events 20 are at less than $3.2^\circ$ from an 
AGN closer than
71~Mpc, while only 6 were expected to be found by chance from an isotropic
distribution of arrival directions. A correlation was first
observed in the data obtained before the end of May 2006, with a very similar
set of parameters, and fixing that set of parameters a priori the subsequent 
data up to August 2007 were studied, confirming the original 
correlation with more than 99\% CL significance in the additional data set
alone. This kind of prescribed test, using an independent data set and a 
priori fixed parameters, is the 
more safe strategy to prevent wrong claims with small statistics. It provides
a clear way of assigning a significance to the observations, without relaying
on penalizations for scanning.

\begin{figure}
  \includegraphics[height=.3\textheight]{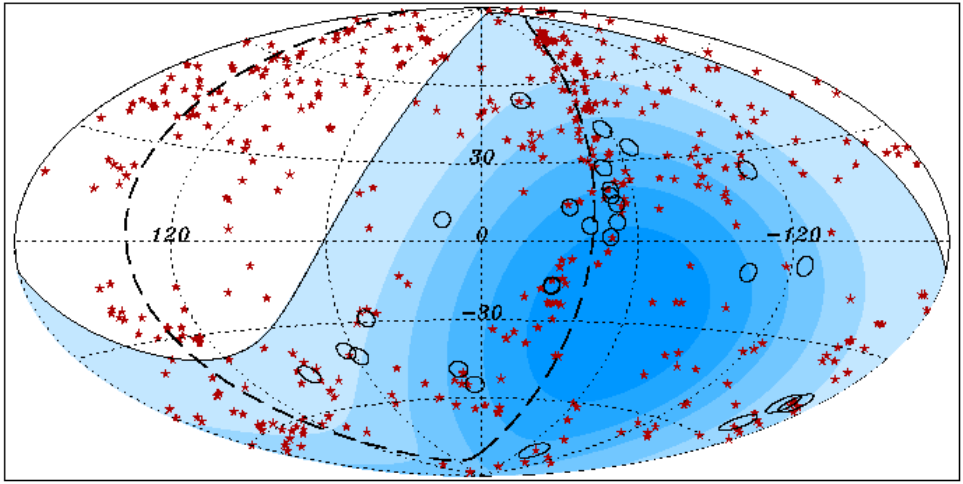}
  \caption{Map in galactic coordinates
    with the positions of the AGNs within 71~Mpc (stars) and the 27 events
    with $E>57$~EeV (circles of 3.2$^\circ$ radius). Shading indicates regions
    of equal exposure \cite{science}.}
\label{skymap}
\end{figure}

The map of the arrival directions and of the AGN positions is shown in
Figure \ref{skymap}. 
A remarkable alignment of several events with the supergalactic plane
(dashed line) is observed, and it is also worth noting that 
two events fall within 3.2$^\circ$ from Centaurus~A, the closest active galaxy.
A further interesting fact is that the energy maximizing the correlation with
AGNs coincides with that maximizing the autocorrelation of the events
themselves \cite{cris08} and is also that for which the spectrum falls to
half of the power law extrapolation from smaller energies \cite{spectrum}.

\paragraph{Log likelihood per event}

\begin{figure}
  \includegraphics[width=3.5in]{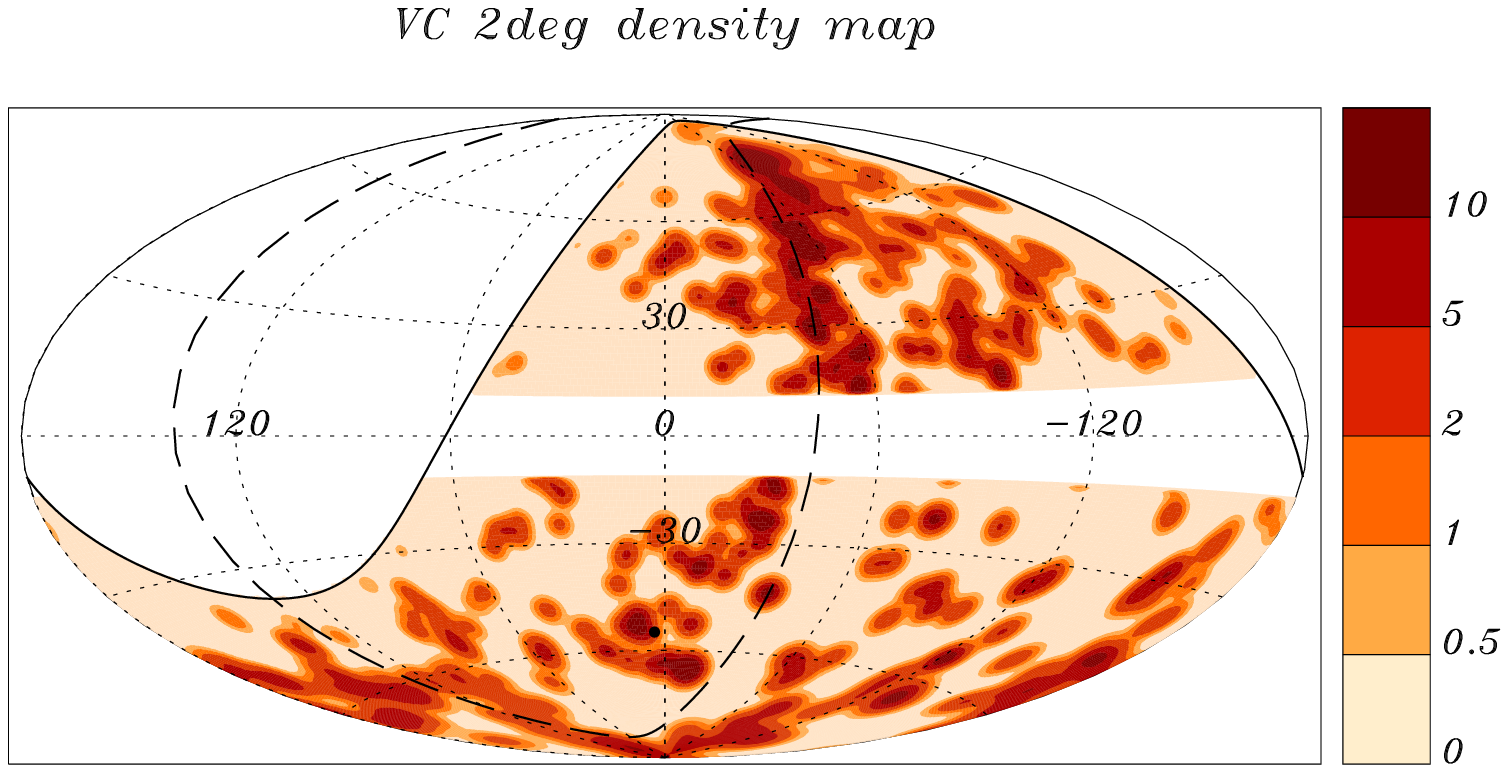}
  \includegraphics[width=1.8in]{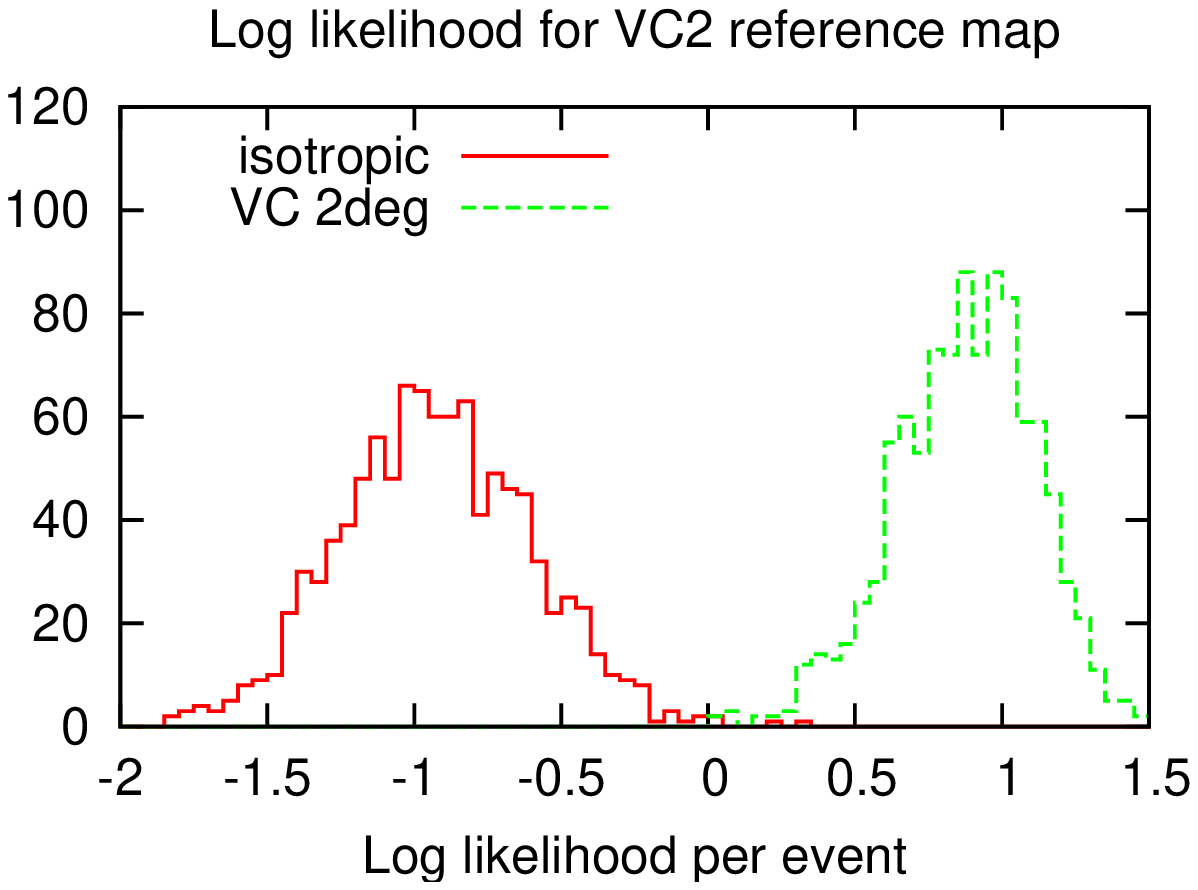}
  \caption{Likelihood map (left panel) and Log likelihood per event
    histograms (right panel)
}
\label{llvc}
\end{figure}

The previous method cannot be applied when the candidate source population 
is large, as the fraction of the sky covered, $p$, becomes of order unity.
A more useful method in this case is to build a probability map for the 
expected arrival directions of events above a given threshold. 
The map can be constructed as follows, a Gaussian of given $\sigma$ is taken
around the direction of each object in the catalog, weighting them by a factor
$$ w(z,E_{th})=\frac{1}{4\pi d_L^2(z) \phi(z)}\int_{E_i(z,E_{th})}^\infty
E^{-s} {\rm d}E. $$
where $d_L$ is the distance to the object, $\Phi(z)$ is the selection 
function of the catalog and the integral term measures the fraction of the 
flux from a source at redshift $z$ that reaches the Earth with an energy 
larger than the threshold $E_{th}$. $E_i$ is the initial energy that the 
particle needs to have at the source to arrive at Earth with $E_{th}$, $s$ is 
the source spectral index. As an example, the left panel of
Figure \ref{llvc} shows the map corresponding to the AGNs in the 
Veron-Cetty and Veron catalog and an energy threshold of 80 EeV.
The likelihood associated to a given set of $N$ observed events is
$L = \prod_{i=1}^N P(\hat u_i),$
with $P(\hat u_i)$ proportional to the map density in each event direction.
In order that the mean value be independent of the total number of events,
it is more convenient to use the log likelihood per event
$$LL =\frac{1}{N} \sum_{i=1}^N \ln P(\hat u_i)$$
The idea is to measure $LL$ of the data using a model reference map. Then
simulate events distributed according to some alternative hypothesis: 
isotropic, following AGNs,... and compute $LL$ for the reference model map.
Then plot the histogram the $LLs$ for each hypothesis: the mean of the 
distribution is independent of the number $N$ of observed events, but the 
width becomes smaller as $N$ grows. When the histograms corresponding to the 
different hypothesis do not overlap the test is good to discriminate 
among them.


In the right panel of Figure \ref{llvc} we show the histograms for 20 
simulated isotropic events and for 20 events following the distribution of AGNs 
in the Veron-Cetty and Veron catalog using a Gaussian window of 2 degrees
size. 

\section{Final remarks}

A variety of different methods are needed to study anisotropies
at different energies and angular scales. It is impossible to review
in a limited space all the interesting ideas that have been put
forward to this scope, I have only presented a (personally biased)
subset of them, as well as a selection of the experimental results
related to the techniques described.

This is a very special time for the field as the CR astronomy is
finally starting and we are getting the first clues on 
the UHECR origin: they are correlated with nearby extragalactic
matter.
There are still many open questions as which are the sources, which is the CR 
composition, how are the relevant magnetic fields?
More data is eagerly waited to clarify these issues.


\begin{theacknowledgments}
I would like to thank my Auger collaborators with whom I learned
a lot in the CR field. I would also like to thank the organizers 
and the attendants of the Cosmic Ray School in the beautiful Arequipa
for a very interesting and lively meeting.
\end{theacknowledgments}

\bibliographystyle{aipproc}

\end{document}